\documentclass[preprint]{aastex}
\usepackage{epsfig}

\shorttitle{Calibrating M dwarf metallicities }
\shortauthors{Woolf \& Wallerstein}

\begin{document}

\title{Calibrating M dwarf metallicities using molecular indices }

\author{Vincent M. Woolf\altaffilmark{1,2} and George
 Wallerstein\altaffilmark{1}}
\email{vwoolf@mail.unomaha.edu, wall@astro.washington.edu}

\altaffiltext{1}{Astronomy Department, University of Washington, Box 351580,
Seattle, WA 98195, USA}
\altaffiltext{2}{Department of Physics, University of Nebraska
at Omaha, 6001 Dodge St., Omaha, NE 68182, USA }

\begin{abstract}
We report progress in the calibration of a method to determine cool dwarf star
metallicities using molecular band strength indices.  The molecular band index
to metallicity relation can be calibrated using chemical abundances calculated
from atomic line equivalent width measurements in high resolution spectra.
Building on previous work, we have measured Fe and Ti abundances in
32 additional M and K dwarf stars to extend the range of temperature and
metallicity covered. A test of our analysis method using warm star -- cool star
binaries shows we can calculate reliable abundances for stars warmer than
3500~K.  We have used abundance measurements for warmer binary or
cluster companions to estimate abundances in 6 additional cool dwarfs. 
Adding stars measured in our previous work and others from the literature
provides 76 stars with Fe abundance and CaH2 and TiO5 index measurements.
The CaH2 molecular index is directly correlated with temperature.  TiO5
depends on temperature and metallicity.  Metallicity can be estimated to within
$\pm0.3$ dex within the bounds of our calibration, which extends from roughly
$\rm [Fe/H] = +0.05$ to $-$1.0 with a limited extension to $-$1.5.

\end{abstract}

\keywords{stars: abundances --- stars: late-type --- stars: subdwarfs}

\section{Introduction}

In \citet{ww05} we reported the measurement of Fe and Ti abundances in 35 M
and K dwarf stars using atomic line equivalent width measurements from
high resolution, $\lambda / \Delta \lambda \approx 33\,000$, spectra.
While the abundance survey provided useful results, it was clear that a 
method of estimating metallicity in cool dwarfs which works for fainter stars
and which required less analysis effort was needed.  Although
low temperature dwarf stars are the most numerous stars in the Galaxy,
their intrinsic faintness means that few of them
are close enough for high resolution spectra to be measured.  Because
the abundances derived depend on the metallicity of the model atmosphere,
several iterations are required for each star before the model and derived
abundances match.

Because low temperature main sequence stars, M dwarf and cooler, make up most
of the baryonic mass of the Galaxy we must know their chemical compositions
if we are to fully understand the chemical composition and evolution of the
Galaxy.  An open problem in modelling the chemical evolution of the Galactic
disk is the `G dwarf problem': fewer metal-poor G dwarfs are observed than
models predict.  The problem has been found to extend to stars with
temperatures as cool as 4700~K \citep{fm97}. With a well-calibrated method
to estimate M dwarf metallicities using low resolution spectra it will be
possible to assemble a stastically significant sample of measurements and
determine if the problem continues to stars with $T=3500$~K or cooler.

\citet{bdu05} combined abundances they measured in 20 binaries with M-dwarf
secondaries and warmer primaries with the metallicity measurements from
\citet{ww05} to calibrate a $\rm M_K$ and $\rm V-K$ vs metallicity relation. 
Because the relation depends on absolute magnitude it will be useful only for
stars close enough for accurate parallaxes to be measured.

In this paper we report a metallicity calibration using the CaH2 and TiO5 
indices.  CaH2 and TiO5 are molecular
indices which measure CaH and TiO band strengths in
cool dwarf stars \citep{rhg95}. These can be measured with lower resolution,
$\lambda / \Delta \lambda \approx 3000$, flux calibrated spectra, and require
only the measurement of relative flux levels in specified wavelength bands in
spectra which have been corrected to zero velocity, which requires much less
observational time and analysis effort than measuring and analyzing equivalent
widths of atomic lines in higher resolution spectra.  This method will
allow metallicities to be estimated for stars at least 3 magnitudes fainter and
considerably more distant than can be observed at high resolution.
With 4-m class telescopes, stars fainter than $\rm V = 16$ can be observed with
reasonable exposure times at this resolution.
For a $\rm M_V = 10$ star this
corresponds to a distance of about 160~pc, a greater distance than that for
which reliable trigonometric parallaxes are available for large numbers of
stars.

We have measured abundances in additional cool dwarfs to extend the range of
temperature and metallicity which can be for which the metallicity relation can
be calibrated. We have also used binaries with F, G, or K
primaries and M dwarf secondaries to test our abundance analysis method and to
extend the temperature and metallicity range.

\section{Observations and reduction}

We obtained spectra for the atomic line abundance analysis using the echelle
spectrograph of the Apache Point Observatory (APO) 3.5-m telescope. The
spectral resolution is $\lambda / \Delta \lambda \approx 33\,000$.  The usable
sections of the spectra cover a range from about 9800 \AA\
to where the measured signal drops off in the blue, normally around 5000 \AA\
for M dwarfs, and well below this for F and G dwarfs.
Because the stars in our first paper included few low metallicity,
$\rm [Fe/H] < -0.5$, stars cooler than 3800 K, we obtained echelle
spectra of additional M dwarfs with molecular band strengths which indicated
they might have low metallicity.

To test the method we use to calculate abundances in M and K dwarfs,
we observed a number of stars in binaries where
one member is an F, G, or early K dwarf and the other is a K or M dwarf with
a temperature in the range covered our other stars.  Most of these were selected
from common proper motion pairs listed by \citet{gc04}. We also observed five
Hyades M and K dwarfs.

We used the Dual Imaging Spectrograph (DIS) of the APO 3.5-m telescope to
measure $\lambda / \Delta \lambda \approx 3000$ spectra of M and K dwarfs for
which no TiO and CaH molecular band indices had been reported.  The red
arm of the spectrograph was set so the spectra cover the range
$\rm 5950 \AA < \lambda < 7650 \AA$.

The echelle spectra were reduced using {\sc iraf} routines as described in 
\citet{ww05}. The DIS spectra were reduced using standard {\sc iraf} routines
to subtract the bias, divide by flat-field spectra, reduce to one-dimensional
spectra, apply HeNeAr lamp spectra wavelength calibration, and do standard
star flux calibrations.

\section{Analysis}
M and K dwarf atmospheric parameters were estimated using V, $\rm K_s$, and
H photometric measurements \citep{2mass,mmh97}  and parallax distances
as described in \citet{ww05}. The magnitudes and parallaxes of the
M and K dwarfs observed for this paper are listed in Table~\ref{T1}.
For several stars we use the Hipparcos parallax of their brighter
binary companion.
Fe and Ti abundances were calculated for the M and K dwarfs using atomic line
equivalent width measurements and {\sc NextGen} \citep{hab99} model
atmospheres as described in \citet{ww05}.  

We measured equivalent widths of Fe~I, Fe~II, Ti~I, and Ti~II lines in the
F, G, and early K stars in binaries with an M or K dwarf using {\sc iraf}.
Temperatures were estimated using
Stromgren photometry \citep{aam96}. Stellar masses and bolometric magnitudes
were estimated using the theoretical $\rm M_V$ vs $\rm B - V$ isochrones
of \citet{b94}.  We calculated stellar radii from the temperatures and
bolometric magnitudes and calculated $\log g$ values from the radii and
masses.  We interpolated Kurucz model atmospheres \citep{k93} to the derived
$T_{\rm eff}$ and $\log g$ values. Microturbulent velocities for the F and G
dwarfs were estimated using the empirical relation:
$\rm \xi = 1.25 + 8\times10^{-4}(T_{eff}-6000) -1.3(\log{\it g} -4.5) km s^{-1}$
\citep{ed93}.  Fe and Ti abundances were calculated from the model atmospheres
and the measured equivalent widths using the current version of the 
LTE stellar analysis program MOOG
\citep{s73}. The exception to this procedure was HIP~13642,
where $T_{\rm eff}$ was estimated from the isochrones because
no Stromgren photometric data were available.

The CaH2 and TiO5 molecular indices are ratios of the average flux levels
in specified wavelength regions. We calculated the indices from the
flux-calibrated DIS spectra using the wavelength regions defined by
\citet{rhg95}.

\section{Results}

\subsection{Chemical abundances}
The atmospheric parameters and the Fe and Ti abundances derived for the 32
cool dwarf stars in Table~\ref{T1} are listed in Table~\ref{T2}. The
metallicity [M/H] parameter listed is the effective metallicity after correcting
for the effect of non-solar $\alpha$-element to Fe abundance ratios as
described in \citet{ww05}. The quoted abundance
uncertainties include the effects of 
uncertainty in temperature, gravity, and microturbulence and the scatter of
abundances determined from different lines in the same star.
The temperature and gravity uncertainties listed in Table~\ref{T2} are the
uncertainties derived from uncertainties in the input parallax and photometry
data and do not include the effects of possible systematic errors, which could 
possibly be as large as about $\pm 100$ to 200 K for the temperature
uncertainty and $\pm 0.3$ to 0.4 for $\log g$. 

To test our method of calculating abundances in cool dwarf stars we observed
binaries with a cool dwarf secondary and a warmer dwarf primary. These
were selected to have large enough angular separations that each component
could be observed individually. Most primaries were F or G dwarfs.  We also
observed five cool dwarfs in the Hyades.  The rationale for this test is that
members of a binary or a cluster which formed from the same material should
have the same chemical composition. Diffusion and nuclear enrichment processes
which can change photospheric abundances should not have any measurable
effect in these unevolved stars. 
The model atmospheres and methods used to find abundances in F and G dwarfs are
well established. The good agreement between solar system meteoritic and solar
photosphere abundances for most elements is evidence that the photospheric
abundances calculated for solar-type stars using these models and methods
are reasonable.  If our method for calculating abundances in cool dwarfs is
accurate then the abundances should agree with those calculated for their
warmer binary or cluster companions.

\begin{figure}
\epsfig{file=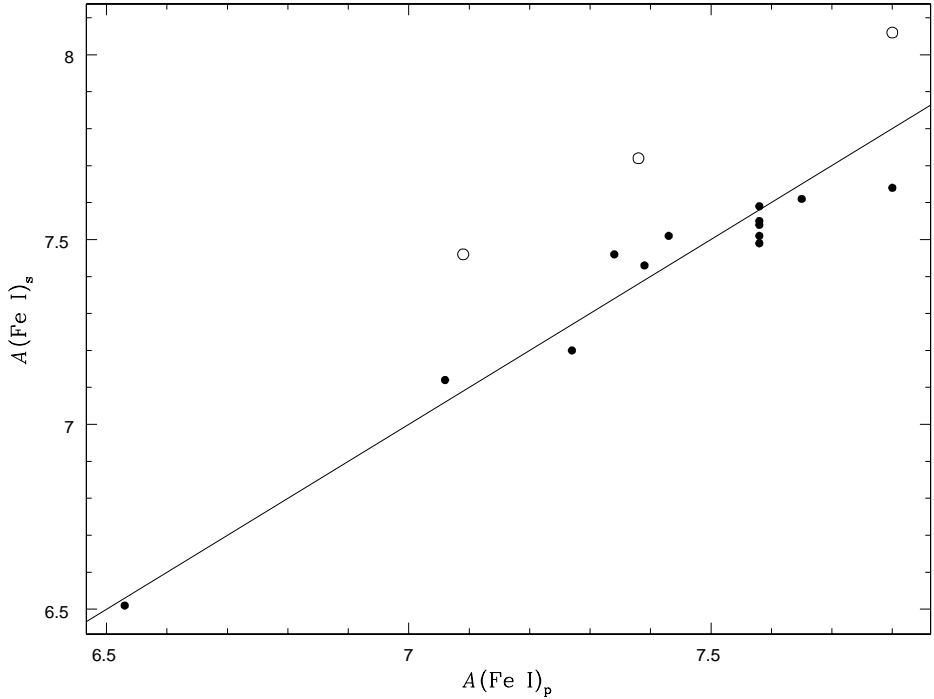, width=10cm, angle=270}
\caption{Cluster or binary warm star vs cool star Fe abundance comparison. Open
circles indicate the secondary stars with $T_{\rm eff}<3500$K. The diagonal line
indicates the location of perfect agreement.}
\label{F1}
\end{figure}

The binary abundances are compared in Table~\ref{T3} and Figure~\ref{F1}.  The
reported F and G dwarf abundance uncertainties include the effects of
uncertainty in temperature, gravity, microturbulence, and the scatter of
abundances determined from different lines.
Temperature uncertainties due to uncertainties in the
Stromgren photometry of the warm stars are less than 40~K. The temperature
uncertainty and the uncertainties in determining the mass and bolometric
magnitude for the warm stars correspond to $\Delta \log g < 0.10$.
We find that the the binary and cluster Fe~I abundances agree to within the 
combined uncertainties except for those with cool dwarfs with
$T_{\rm eff} < 3500$~K, which are indicated by open circles in the figure.
The same pattern is seen for the Ti~I abundances except that HD~18143B,
BD+19~1185B, BD+17~791C, and LP~13-691 have abundance estimates a bit 
smaller than their warmer companions, even allowing for the uncertainties.
Our method of measuring chemical abundances in cool dwarfs appears to 
provide accurate results for stars with temperatures greater than 3500~K.
A weakness of this test is that we were unable to observe only one such
binary with $\rm [Fe/H] < -0.5$: low metallicity stars are less common, and
identifying low metallicty cool dwarfs which are also in widely separated
visual binaries with F or G dwarf companions is even more difficult.

The fact that we calculate similar abundances for both members of the binaries
where the cooler member is warmer than 3500~K implies that the abundances
we find for the warmer stars are not significantly affected by the possible
problems caused by molecular bands. The metallicities of the binaries we
studied were close to the solar metallicity. Stars of lower metallicity would
be less affected by weak line blanketting.

There appears to be a systematic error causing the abundances calculated for
stars cooler than 3500 K to be too large.
It may be that the model atmospheres used do not sufficiently model
the $\rm H_2O$ opacity which starts becoming much stronger at temperatures
cooler than about 3500 K.  It is also possible that increasing molecular
band strengths at the cooler temperatures depress the apparent continuum in
a molecular line haze, leading us to overestimate the atomic line equivalent
widths.  We note that because of this result, we are no longer certain of the
abundances reported for LHS~450 in \citet{ww05}, the only $T_{\rm eff}<3500$~K
star in that paper.

It should be possible, however, to calibrate our molecular index vs
metallicity correlation to lower temperatures by including visual binaries. 
The last three
binaries in Table~\ref{T3} have secondaries which are too cool for
abundances to be measured.  We will assume that they have the abundances of
their warmer companions.

\subsection{Molecular index - metallicity calibration}

The molecular indices and Fe abundances to be used for the metallicity
calibration are listed in Table~\ref{T4}.  The stars listed include all those
from this paper and \citet{ww05} for which molecular index data are available
and 12 more from the other published reports.  The abundances used for cool
dwarfs in binaries with an F, G, or early K star are those of their warmer star.
Abundances for stars in the Hyades or in a binary with two cool stars warmer
than 3500~K are given by the average cluster or binary
abundance.  The uncertainties of the CaH2 and TiO5 index measurements are
about $\pm 0.04$ \citep{rhg95} or $\pm 5$ to $10\%$ \citep{zm04}.

\begin{figure}
\epsfig{file=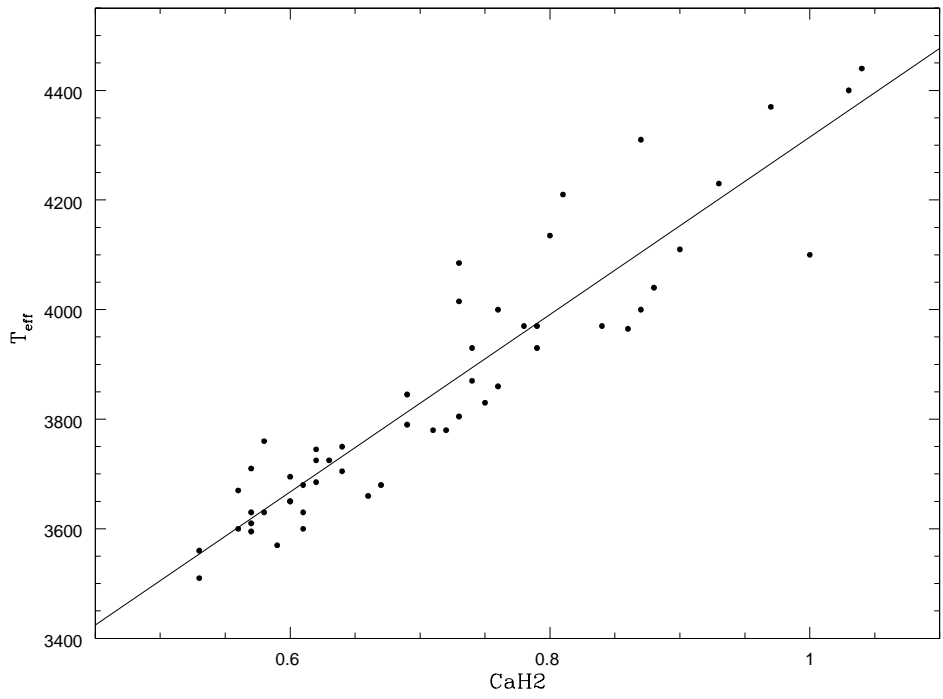, width=10cm, angle=270}
\caption{Temperature vs CaH2 index for program stars.  The line is a
least squares fit: $\rm T_{eff} = (2696 + 1618\times CaH2)$~K }
\label{F2}
\end{figure}

\begin{figure}
\epsfig{file=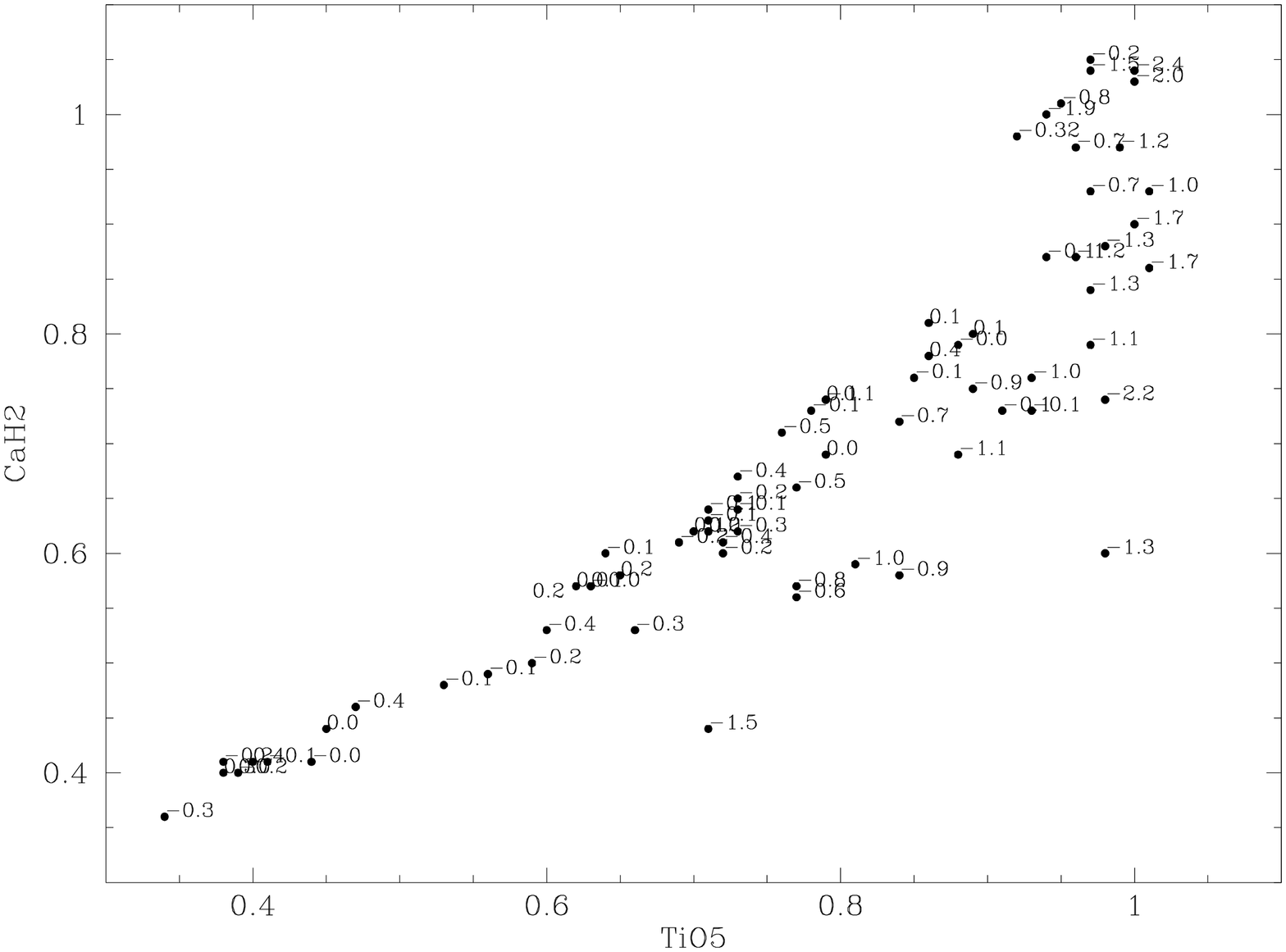, width=10cm, angle=270}
\caption{CaH2 vs TiO5.  Numbers next to points indicate the [Fe/H] values
for each star. }
\label{F3}
\end{figure}

CaH2 is well correlated with effective temperature, as shown in Figure~\ref{F2}.
TiO5 depends on temperature and metallicity.  
The locations of the stars in the CaH2 vs TiO5 plane and their [Fe/H] abundances
are shown in Figure~\ref{F3}.  Metallicity decreases to the lower right in the
figure as expected: for a given temperature or CaH2 value, a smaller TiO5 value
indicates a smaller metallicity.

\begin{figure}
\epsfig{file=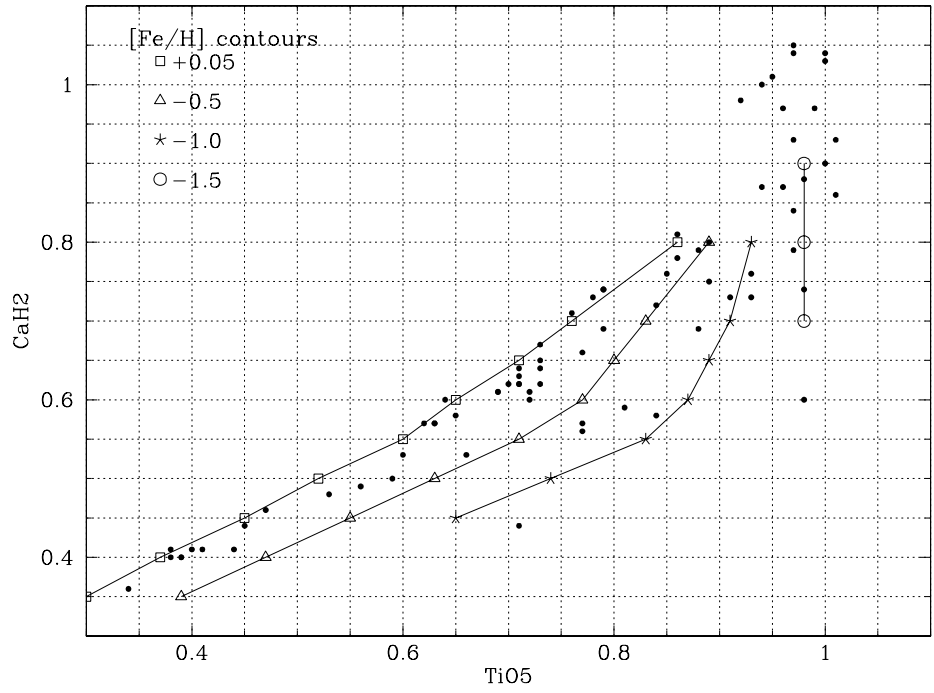, width=10cm, angle=270}
\caption{Equal-metallicity contours in CaH2 vs TiO5 }
\label{F4}
\end{figure}

We were unable to find an empirical polynomial fit to the data which corresponds
well to the data.  We have estimated the locations of equal-metallicity lines
in CaH2 vs TiO5 by eye, as shown in Table~\ref{T5} and Figure~\ref{F4}.
Because the molecular band strengths decrease at higher temperatures, the fits
start to converge for $\rm CaH2 \gtrsim 0.8$ or $T_{\rm eff} \gtrsim 4000$ K.  
Molecular band strengths are poor indicators of metallicity at temperatures 
where the bands are very weak and the molecular index uncertainties correspond
to large metallicity uncertainty.  The $\pm 0.04$  molecular index uncertainties
correspond to an [Fe/H] uncertainty of about $\pm 0.3$ through the entire region
covered by our calibration grid.

\section{Discussion}
We have now determined the Fe and Ti abundances for 84 M and K dwarf stars
using high resolution spectra and equivalent width abundance analysis.
We have tested our analysis method using stars in binaries and a cluster 
and find that it appears to give reliable abundances for stars warmer than
3500~K. 

When we include abundance data for 12 stars from other researchers we have
76 stars with measured Fe abundances and CaH2 and TiO5 indices.  We have used
these to create a rough molecular index -- metallicity calibration.

The main shortcoming of our data is that we have few  low metallicity stars
with $T_{\rm  eff} < 4000$~K.  Our data are not yet sufficient to determine 
whether the difficulty in identifying very low metallicity stars is partly
caused by an `M dwarf problem' similar to the G dwarf problem, where low
metallicity stars are less common than predicted by Galactic star formation 
and chemical evolution models.  At a given temperature, low metallicity stars
are fainter than solar metallicity stars; they are ``subdwarfs''. This means
they must be physically closer to appear bright enough for a spectrum with
sufficient signal to be obtained.

We have been granted time on the Hobby-Eberly Telescope to observe several M
dwarf stars with CaH2 and TiO5 indices which indicate they have
$\rm [Fe/H] < -1.0$, but which are too faint to observe with APO.  The
abundances calculated from these spectra will help populate the low
metallicity region of the CaH2 - TiO5 plane.

When the calibration is adequately defined it will be possible to estimate
the metallicities of thousands of cool dwarf stars. Thousands of spectra of
red dwarf stars have already been observed in the Sloan Digital Sky
Survey.  While the abundances estimated through a molecular index calibration
will necessarily have larger uncertainties than those calculated through the
analysis of atomic line strengths in high resolution spectra, they will be
sufficient to allow statistical study of the relative numbers of cool dwarfs
of different metallicities and a determination whether the G dwarf problem
continues to lower masses, i.e. whether low metallicity M dwarfs are more scarce
than models predict.

\acknowledgments

We thank Nicole Silvestri for obtaining some of the DIS spectra used for
our molecular index measurements,
Peter Hauschildt for continuing help with NextGen model atmospheres
and  Suzanne Hawley for helpful discussions about low mass subdwarfs. We thank
Iain Reid for providing low resolution spectra to recheck the molecular indices
of a couple stars.
This research has made use of the SIMBAD database, operated at CDS, Strasbourg,
France. This research has made use of NASA's Astrophysics Data System
Bibliographic Services. 
The authors gratefully acknowledge the financial support of the
Kennilworth Fund of the New York Community Trust.

Facilities: \facility{APO}

\begin{table*}
\begin{minipage}[l]{100mm}
\footnotesize
\setlength{\tabcolsep}{0.05in}
\caption{M and K dwarf magnitudes and parallaxes}\label{T1}
\begin{tabular}{lllrrrcrcrrc}
\hline \hline
Star &Alternate & Spectral& V & $\pm$\footnote{In the absence of a reported V
uncertainty estimate we used 0.04}& K$\rm _s$ & $\pm$& H & $\pm$& $\pi$ &
$\pm$& $\pi$ \cr
 & name & type\footnote{taken from Gizis (1997), SIMBAD, Lee (1984), and Luyten
(1979)}     &&&&  &          &   &(mas)&(mas)&source\footnote{Hip: ESA (1997),
Yal: van Altena, Lee, \& Hoffleit (1995), Hya: Hyades distance,
Nav: Harrington \& Dahn (1980)}\cr

HIP 1386&GJ 3023&M2&11.518&0.020&7.241&0.011&7.499&0.023&42.65&2.77&Hip \\
HIP 17743&LHS 1594&M1&11.046&0.015&7.110&0.018&7.382&0.057&57.59&2.56&Hip\\
HIP 26801&HD 233153&M0.5&9.79&0.03&5.759&0.016&5.963&0.016&81.17&0.53&Hip\\
HIP 37798&GJ 287&K5&10.193&0.020&6.768&0.021&6.992&0.043&40.58&2.18&Hip \\
HIP 59514&HD 238090&M0&9.79&0.03&6.059&0.017&6.245&0.017&65.29&1.47&Hip\\
HIP 67308&GJ 1177A&K4&8.94&0.04&5.557&0.017&5.725&0.021&61.07&2.93&Hip\\
HIP 86087&GJ 685&M0.5&9.97&0.01&6.066&0.018&6.271&0.017&70.95&1.09&Hip \\
HIP 89490&HD 348274&M0&10.840&0.019&6.964&0.017&7.172&0.022&43.10&2.18&Hip \\
HIP 98906&LHS 482&sdM1.5&11.97&0.04&8.113&0.015&8.364&0.015&63.23&6.44&Hip\\
HIP 105932&GJ 828.2&M0.5 & 11.097&0.013&7.166&0.022&7.458&0.053&61.57&2.61&Hip\\
HIP 117383&GJ 907&M1&12.060&0.033&7.933&0.021&8.142&0.037&60.56&3.24&Hip\\
HD 7895B & LHS 1229&K7&10.705&0.029&7.190&0.013&7.369&0.025&36.16&1.00&Hip\\
HD 11964B & GJ 81.1B&K7&11.211&0.013&7.597&0.027&7.763&0.021&29.43&0.91&Hip \\
HD 18143B&GJ 118.2B&K7&9.80&0.04&6.170&0.036&6.299&0.038&43.71&1.26&Hip\\
HD 263175B&GJ 3409B&M0.5&12.17&0.01&8.184&0.014&8.428&0.031&40.02&1.22&Hip\\
HD 285804&\nodata&K5&11.098&0.022&7.758&0.021&7.994&0.063&21.5&5.0&Hya\\
BD-1 293B&\nodata&K&10.52&0.04&7.390&0.021&7.540&0.029&27.04&0.86&Hip \\
BD+17 719C&\nodata&\nodata&11.30&0.02&7.977&0.018&8.160&0.017&21.5&5.0&Hya\\
BD+19 1185B&LHS 1812&M1&13.40&0.04&9.822&0.028&10.011&0.021&14.86&2.50&Hip\\
BD+23 2207B&GJ 387B&M1&11.40&0.04&7.593&0.026&7.794&0.047&44.01&0.75&Hip\\
BD+24 4B&G 130-47&\nodata&11.480&0.035&8.087&0.027&8.268&0.024&22.07&2.31&Hip\\
GJ 107B&BD+48 746B&M1.5&9.87&0.04&5.865&0.021&6.080&0.038&89.03&0.79&Hip\\
GJ 129&LHS 169&esdK7&14.16&0.04&10.819&0.018&11.012&0.023&30.0&2.3&Yal\\
GJ 1177B&\nodata&K5&9.12&0.04&5.642&0.027&5.799&0.024&61.07&2.93&Hip\\
GJ 3212&\nodata &M0.5&11.630&0.015&7.681&0.020&7.911&0.034&\nodata&\nodata&\nodata \\
GJ 3278 &\nodata&M0.5&12.546&0.029&8.561&0.016&8.787&0.019&21.5&5.0&Hya\\
GJ 3290&\nodata&M1.5&13.055&0.014&8.826&0.024&9.067&0.028&21.5&5.0&Hya\\
GJ 3825&LHS 364&esdM1.5&14.551&0.027&10.860&0.017&11.015&0.016&35.&5.&Nav\\
GJ 9722&LHS 64&sdM1.5&13.30&0.04&9.390&0.016&9.583&0.029&41.8&2.7&Yal\\
LHS 491&G 210-19&sdM1.5&14.74&0.03&10.869&0.024&11.102&0.027&20.4&4.2&Yal\\
LP 13-691&\nodata&M0&11.84&0.04&8.224&0.018&8.390&0.018&21.5&5.0&Hya \\
2MASS 2203769-&HIP 108923B&\nodata&12.64&0.04&8.865&0.020&9.062&0.020&19.61&1.57&Hip\\
\hspace{1cm}2452313
\end{tabular}
\end{minipage}
\end{table*}

\begin{table*}
\begin{minipage}[l]{150mm}
\caption{M and K dwarf parameters and abundances}\label{T2}
\begin{tabular}{lcccrrr}
\hline \hline
Star & T$_{\rm eff}$ & $\log g$\footnote{we used $\log g = 5.0 \pm 0.5$
for the three stars where parallax was unavailable.}  &
$\xi$&[M/H]& [Fe/H] & [Ti/H]\footnote{we use
$A{\rm (Fe)}_\odot = 7.45$, $A{\rm (Ti)}_\odot = 5.02$ }  \\
        & K  && km s$^{-1}$ &\\
\hline
HIP 1386   &$3600\pm60$&$4.67\pm0.11$& 1.0&   0.15&$0.16\pm0.10$&$0.13\pm0.09$\\
HIP 17743&$3685\pm35$&$4.84\pm0.07$& 1.0&$-$0.25&$-0.30\pm0.07$&$-0.22\pm0.07$\\
HIP 26801  &$3725\pm20$&$4.71\pm0.03$& 1.0&  0.15&$0.16\pm0.09$&$0.09\pm0.10$ \\
HIP 37798&$4135\pm45$&$4.67\pm0.09$& 1.0& 0.10&$0.10\pm0.09$&$0.11\pm0.11$  \\
HIP 59514&$3845\pm25$&$4.69\pm0.05$& 1.0&$-$0.03&$-0.05\pm0.08$&$-0.08\pm0.09$\\
HIP 67308&$4085\pm40$&$4.57\pm0.08$& 1.4&$-$0.12&$-0.16\pm0.10$&$-0.26\pm0.11$\\
HIP 86087&$3750\pm15$&$4.71\pm0.03$& 1.0& 0.02&$0.01\pm0.08$&$0.01\pm0.10$  \\
HIP 89490&$3660\pm20$&$4.56\pm0.07$& 1.8&$-$0.44&$-0.53\pm0.08$&$-0.53\pm0.10$\\
HIP 98906&$3670\pm35$&$5.03\pm0.18$& 1.0&$-$0.52&$-0.62\pm0.10$&$-0.22\pm0.09$\\
HIP 105932&$3680\pm40$&$4.87\pm0.09$&0.2&$-$0.30&$-0.37\pm0.05$&$-0.18\pm0.04$\\
HIP 117383&$3560\pm25$&$4.98\pm0.09$&1.0&$-$0.28&$-0.33\pm0.05$&$-0.38\pm0.08$\\
HD 7895B &$4000\pm30$&$4.69\pm0.06$&1.0&$-$0.04&$-0.07\pm0.08$&$-0.17\pm0.10$ \\
HD 11964B &$3930\pm25$&$4.67\pm0.06$&1.0&0.00&$-0.02\pm0.09$&$0.02\pm0.11$  \\
HD 18143B&$3970\pm45$&$4.52\pm0.07$&1.0&0.18&$0.19\pm0.11$&$0.07\pm0.13$  \\
HD 263175B&$3655\pm20$&$4.89\pm0.06$&1.0&$-$0.20&$-0.25\pm0.07$&$-0.25\pm0.08$\\
HD 285804&$4210\pm60$&$4.57\pm0.28$&1.0&0.10&$0.10\pm0.09$&$0.07\pm0.13$  \\
BD-1 293B&$4310\pm45$&$4.63\pm0.07$&1.75&$-$0.06&$-0.09\pm0.08$&$-0.20\pm0.08$\\
BD+17 719C&$4185\pm30$&$4.65\pm0.27$&1.0&0.05&$0.04\pm0.09$&$-0.05\pm0.11$\\
BD+19 1185B&$3820\pm40$&$4.77\pm0.23$&1.0&$-$0.78&$-0.94\pm0.07$&$-1.00\pm0.08$ \\
BD+23 2207B&$3740\pm40$&$4.83\pm0.05$&1.0&$-$0.27&$-0.33\pm0.06$&$-0.28\pm0.08$ \\
BD+24 4B&$4085\pm40$&$4.67\pm0.04$&1.0&$-$0.11&$-0.15\pm0.09$&$-0.11\pm0.10$ \\
GJ 107B&$3715\pm20$&$4.77\pm0.03$&1.0&0.07&$0.06\pm0.08$&$0.06\pm0.09$ \\
GJ 129&$3965\pm40$&$5.0\pm0.5$&1.0&$-$1.33&$-1.66\pm0.05$&$-1.34\pm0.15$  \\
GJ 1177B&$4015\pm40$&$4.57\pm0.09$&1.2&$-$0.06&$-0.09\pm0.10$&$-0.20\pm0.11$ \\
GJ 3212&$3705\pm25$&$5.0\pm0.5$&0.0&$-$0.06&$-0.08\pm0.05$&$0.01\pm0.05$  \\
GJ 3278 &$3745\pm20$&$4.69\pm0.27$&1.0&0.14&$0.14\pm0.10$&$0.12\pm0.11$  \\
GJ 3290&$3630\pm15$&$4.75\pm0.29$&1.0&0.10&$0.10\pm0.11$&$0.02\pm0.13$ \\
GJ 3825&$3695\pm25$&$5.0\pm0.5$&1.5&$-$1.09&$-1.34\pm0.10$&$-1.17\pm0.13$ \\
GJ 9722&$3595\pm30$&$5.08\pm0.12$&1.0&$-$0.70&$-0.83\pm0.04$&$-0.75\pm0.07$  \\
LHS 491&$3630\pm30$&$5.08\pm0.32$&1.0&$-$0.78&$-0.93\pm0.08$&$-0.68\pm0.07$  \\
LP 13-691&$3955\pm35$&$4.65\pm0.27$&1.20&0.07&$0.06\pm0.11$&$-0.06\pm0.13$ \\
2MASS 2203769-&$3805\pm35$&$4.72\pm0.12$&1.0&$-$0.09&$-0.12\pm0.09$&$-0.17\pm0.10$  \\
\hspace{1cm}2452313
\end{tabular}
\end{minipage}
\end{table*}

\begin{deluxetable}{lcccccccccccccll}
\tabletypesize{\scriptsize}
\setlength{\tabcolsep}{0.05in}
\rotate
\tablewidth{8.5in}
\tablecaption{
Binary and cluster abundances\label{T3}}
\tablehead{
Primary&T$_{\rm eff}$ & $\log g$& $\xi$& Fe I\tablenotemark{a}
&n&Fe II&n&Ti I&n&Ti II&n&
Secondary&T& Fe I& Ti I \\
&K&& km s$^{-1}$ &&lines&&lines&&lines&&lines&&K
}
\startdata
HIP 9094&5140&3.86&1.39&$7.39\pm0.05$&45&$7.55\pm0.11$&16&$4.86\pm0.09$&15&$4.87\pm0.13$&3&HD 11964B&3930&$7.43\pm0.09$&$5.04\pm0.11$\\
HIP 12777&6200&4.25&1.60&$7.43\pm0.06$&29&$7.51\pm0.09$&7&$4.98\pm0.10$&4&$4.86\pm0.10$&2&GJ 107B&3710&$7.51\pm0.08$&$5.08\pm0.09$\\
HIP 13642&5150&4.52&0.50&$7.80\pm0.08$&33&$7.82\pm0.10$&17&$5.54\pm0.10$&34&$5.31\pm0.13$&2&HD 18143B&3970&$7.64\pm0.11$&$5.09\pm0.13$\\
HIP 13642&\nodata&\nodata&\nodata&\nodata&\nodata&\nodata&\nodata&\nodata&\nodata&\nodata&\nodata&HD 18143C&3110&8.06&5.83\\
HIP 14286&5570&4.37&1.08&$7.09\pm0.07$&29&$7.10\pm0.11$&13&$4.88\pm0.09$&16&$4.80\pm0.17$&5&LHS 1494&3410&7.46&5.15\\
HIP 26779&5135&4.54&0.50&$7.65\pm0.06$&26&$7.81\pm0.10$&10&$5.16\pm0.09$&16&$5.41\pm0.18$&4&HIP 26801&3725&$7.61\pm0.09$&$5.11\pm0.10$\\
HIP 28671&5500&4.42&0.94&$6.53\pm0.08$&34&$6.38\pm0.12$&8&$4.24\pm0.12$&5&$4.11\pm0.14$&2&BD+19 1185B&3820&$6.51\pm0.07$&$4.02\pm0.08$\\
HIP 32423&4830&4.71&0.75&$7.27\pm0.06$&29&$7.40\pm0.17$&6&$4.86\pm0.09$&17&$4.98\pm0.11$&3&HD 263175B&3650&$7.20\pm0.07$&$4.77\pm0.08$\\
HIP 50384&6030&4.27&1.57&$7.06\pm0.05$&24&$7.08\pm0.10$&13&$4.68\pm0.09$&6&$4.68\pm0.10$&4&BD+23 2207B&3740&$7.12\pm0.06$&$4.74\pm0.08$\\
HIP 67308&4085&4.57&1.4&$7.29\pm0.10$&15&\nodata&\nodata&$4.76\pm0.11$&34&\nodata&\nodata&GJ 1177B&4015&$7.36\pm0.10$&$4.82\pm0.11$ \\
HIP 86036&5710&4.38&1.17&$7.34\pm0.05$&27&$7.63\pm0.10$&8&$4.99\pm0.07$&13&$5.06\pm0.13$&4&HIP 86087&3750&$7.46\pm0.08$&$5.03\pm0.10$\\
HIP 114156&4250&4.69&1.0&$7.38\pm0.08$&24&\nodata&\nodata&$4.93\pm0.10$&39&\nodata&\nodata&G 275-2&3225&7.72&5.28\\
Hyades\tablenotemark{b}&\nodata&\nodata&\nodata&$7.58\pm0.01$&\nodata&\nodata&\nodata&$5.18\pm0.01$&\nodata&\nodata&\nodata&BD+17 719C&4185&$7.49\pm0.09$&$4.97\pm0.11$\\
Hyades&\nodata&\nodata&\nodata&$7.58\pm0.01$&\nodata&\nodata&\nodata&$5.18\pm0.01$&\nodata&\nodata&\nodata&GJ 3290&3630&$7.55\pm0.11$&$5.04\pm0.13$\\
Hyades&\nodata&\nodata&\nodata&$7.58\pm0.01$&\nodata&\nodata&\nodata&$5.18\pm0.01$&\nodata&\nodata&\nodata&GJ 3278&3745&$7.59\pm0.10$&$4.14\pm0.11$\\
Hyades&\nodata&\nodata&\nodata&$7.58\pm0.01$&\nodata&\nodata&\nodata&$5.18\pm0.01$&\nodata&\nodata&\nodata&LP 13-691&3955&$7.51\pm0.11$&$4.96\pm0.13$\\
Hyades&\nodata&\nodata&\nodata&$7.58\pm0.01$&\nodata&\nodata&\nodata&$5.18\pm0.01$&\nodata&\nodata&\nodata&HD 285804&4210&$7.55\pm0.09$&$5.09\pm0.13$ \\
HIP 75069&5185&4.64&0.50&$7.21\pm0.04$&27&$7.30\pm0.09$&10&$4.79\pm.07$&14&$4.80\pm0.14$&4&LTT 14560&\nodata&\nodata&\nodata \\
HIP 97675&6010&4.05&1.84&$7.46\pm0.06$&28&$7.40\pm0.11$&12&$4.93\pm0.09$&5&$4.88\pm0.12$&5&GJ 768.1B&\nodata&\nodata&\nodata \\
HIP 99452&5250&4.55&0.58&$7.48\pm0.07$&25&$7.42\pm0.07$&8&$5.32\pm0.09$&15&$5.11\pm0.17$&4&GJ 283.2B&\nodata&\nodata&\nodata \\
\enddata
\tablenotetext{a}{Abundances in this table are reported as
$A(X)\equiv \log(N_X/N_H) +12.00$}
\tablenotetext{b}{Hyades abundances from \citet{psc03}}
\end{deluxetable}

\begin{deluxetable}{lcrcccc}
\tabletypesize{\scriptsize}
\tablecaption{Abundances and molecular indices for calibration\label{T4}}
\tablehead{
Star & T$_{\rm eff}$ & [Fe/H] &CaH2&TiO5&Fe source\tablenotemark{a}
&Index source\tablenotemark{b}
}
\startdata
HIP 1386&3600&0.16&0.56&0.59&1&1  \\
HIP 17743&3685&$-$0.30&0.62&0.73&1&3 \\
HIP 26801&3725&0.20&0.62&0.71&2&3\\
HIP 27928&4370&$-$0.73&0.97&0.96&3&1\\
HIP 37798&4135&0.10&0.80&0.89&1&3 \\
HIP 59514&3845&$-$0.05&0.69&0.79&1&3  \\
HIP 67308&4085&$-$0.13&0.73&0.93&2&1\\
HIP 86087&3750&$-$0.11&0.64&0.71&2&3\\
HIP 89490&3660&$-$0.53&0.66&0.77&1&3\\
HIP 98906&3670&$-$0.62&0.56&0.77&1&2\\
HIP 105932&3680&$-$0.37&0.67&0.73&1&3\\
HIP 117383&3560&$-$0.33&0.53&0.66&1&1\\
HD 7895B &4000&$-$0.07&0.76&0.85&1&3\\
HD 11964B &3930&$-$0.06&0.74&0.79&2&3\\
HD 18143B&3970&0.35&0.78&0.86&2&3\\
HD 18143C&\nodata&0.35&0.41&0.40&2&3\\
HD 33793&3570&$-$0.99&0.59&0.81&3&2\\
HD 36395&3760&0.21&0.58&0.65&3&3 \\
HD 88230&3970&$-$0.03&0.79&0.88&3&3\\
HD 95735&3510&$-$0.42&0.53&0.60&3&3\\
HD 97101B&3610&0.02&0.57&0.62&3&3\\
HD 119850&3650&$-$0.10&0.60&0.64&3&3\\
HD 178126&4530&$-$0.72&0.93&0.97&3&1\\
HD 199305&3720&$-$0.13&0.63&0.71&3&3\\
HD 217987&3680&$-$0.22&0.61&0.69&3&1\\
HD 263175B&3655&$-$0.18&0.60&0.72&2&3\\
HD 285804&4210&0.13&0.81&0.86&2&1\\
BD-1 293B&4310&$-$0.09&0.87&0.94&1&1\\
GJ 81.1B&\nodata&0.09&0.74&0.79&4&3\\
GJ 105B&\nodata&$-$0.19&0.41&0.38&4&3\\
GJ 107B&3715&$-$0.02&0.57&0.63&2&3\\
GJ 129&3965&$-$1.66&0.86&1.01&1&2\\
GJ 166C&\nodata&$-$0.33&0.36&0.34&4&3\\
GJ 212&\nodata&0.04&0.62&0.71&4&3\\
GJ 231.1B&\nodata&$-$0.02&0.41&0.44&4&3\\
GJ 250B&\nodata&$-$0.15&0.50&0.59&4&3\\
GJ 283.2B&\nodata&0.03&0.40&0.39&2&3\\
GJ 297.2B&\nodata&$-$0.09&0.49&0.56&4&3\\
GJ 324B&\nodata&0.32&0.40&0.38&4&3\\
GJ 505B&\nodata&$-$0.25&0.65&0.73&4&3\\
GJ 768.1B&\nodata&0.01&0.44&0.45&2&3\\
GJ 783.2B&\nodata&$-$0.16&0.40&0.39&4&3\\
GJ 797B&\nodata&$-$0.07&0.48&0.53&4&3\\
GJ 1177B&4015&$-$0.13&0.70&0.91&2&1\\
GJ 3212&3705&$-$0.08&0.64&0.73&1&3\\
GJ 3278 &3745&0.14&0.62&0.70&1&3\\
GJ 3290&3630&0.13&0.57&0.63&2&3\\
GJ 3825&3695&$-$1.34&0.60&0.98&1&2\\
GJ 9722&3595&$-$0.83&0.57&0.77&1&2\\
LHS 12&3830&$-$0.89&0.75&0.89&3&2\\
LHS 38&3600&$-$0.43&0.61&0.72&3&3\\
LHS 42&3860&$-$1.05&0.76&0.93&3&2\\
LHS 104&3970&$-$1.33&0.84&0.97&3&2\\
LHS 170&4230&$-$0.97&0.93&1.01&3&2\\
LHS 173&4000&$-$1.19&0.87&0.96&3&2\\
LHS 174&3790&$-$1.11&0.69&0.88&3&2\\
LHS 182&3870&$-$2.15&0.74&0.98&3&2\\
LHS 236&4040&$-$1.32&0.88&0.98&3&2\\
LHS 343&4110&$-$1.74&0.90&1.00&3&2\\
LHS 467&3930&$-$1.10&0.79&0.97&3&2\\
LHS 491&3630&$-$0.93&0.58&0.84&1&2\\
LHS 541&\nodata&$-$1.54&0.44&0.71&5&3\\
LHS 1138&4620&$-$2.39&1.04&1.00&3&1\\
LHS 1482&4100&$-$1.88&1.00&0.94&3&1\\
LHS 1494&\nodata&$-$0.36&0.46&0.48&2&3\\
LHS 1819&4670&$-$0.77&1.01&0.95&3&1\\
LHS 1841&4440&$-$1.47&1.04&0.97&3&1\\
LHS 2161&4500&$-$0.32&0.98&0.92&3&1\\
LHS 2715&4590&$-$1.16&0.97&0.99&3&1\\
LHS 3084&3780&$-$0.73&0.72&0.84&3&2\\
LHS 3356&3630&$-$0.20&0.61&0.69&3&3\\
LHS 5337&3780&$-$0.50&0.71&0.76&3&1\\
G 39-36&4400&$-$2.00&1.03&1.00&3&1\\
G 275-2&\nodata&$-$0.07&0.41&0.41&2&3\\
2MASS 2203769-2452313&3805&$-$0.12&0.73&0.78&1&1\\
LTT 14560&\nodata&$-$0.24&1.05&0.97&2&1\\
\enddata
\tablenotetext{a}{1: this paper, 2: this paper: binary or cluster abundance, 3:
\citet{ww05}, 4: \cite{bdu05}, 5: \cite{f00}}
\tablenotetext{b}{1: this paper, 2: \citet{g97}, 3: \citet{rhg95} and
\cite{hgr96} }
\end{deluxetable}

\begin{table*}
\caption{CaH2, TiO5, [Fe/H] grid points}\label{T5}
\begin{tabular}{cccccccc}
\hline \hline
\multicolumn{2}{c}{[Fe/H]=0.05}&\multicolumn{2}{c}{[Fe/H]=$-$0.50}&
\multicolumn{2}{c}{[Fe/H]=$-$1.0}&\multicolumn{2}{c}{[Fe/H]=$-$1.5} \\
CaH2 & TiO5& CaH2&TiO5&CaH2&TiO5&CaH2&TiO5\\
\hline
0.35  & 0.30  & 0.35  & 0.39  & \nodata&\nodata&\nodata&\nodata\\
0.40  & 0.37  & 0.40  & 0.47  & \nodata&\nodata&\nodata&\nodata\\
0.45  & 0.45  & 0.45  & 0.55  & 0.45  & 0.65  &\nodata&\nodata \\
0.50  & 0.52  & 0.50  & 0.63  & 0.50  & 0.74  &\nodata&\nodata \\
0.55  & 0.60  & 0.55  & 0.71  & 0.55  & 0.83  &\nodata&\nodata \\
0.60  & 0.65  & 0.60  & 0.77  & 0.60  & 0.87  &\nodata&\nodata \\
0.65  & 0.71  & 0.65  & 0.80  & 0.65  & 0.89  &\nodata&\nodata \\
0.70  & 0.76  & 0.70  & 0.83  & 0.70  & 0.91  & 0.70  & 0.98 \\
0.80  & 0.86  & 0.80  & 0.89  & 0.80  & 0.93  & 0.80  & 0.98 \\
\end{tabular}
\end{table*}

\end{document}